%% file: example_paper.tex
\documentclass{article}
\usepackage[x11names]{xcolor}  

\usepackage{graphicx}  
\usepackage{subcaption} 

\usepackage{microtype}
\usepackage{booktabs} 
\usepackage{rotating}
\usepackage{multirow}
\usepackage{tcolorbox}

\usepackage{amsmath,amsfonts,amssymb,amsthm,bbm,bm, braket}
\usepackage{tensor}
\usepackage{tikz}
\usepackage{tikz-network}
\usetikzlibrary{patterns,decorations.pathreplacing,math}
\usepackage{xcolor}
\usepackage{color}
\definecolor{myred}{RGB}{232,102,102}
\definecolor{myblue}{RGB}{187,187,255}
\definecolor{myviolet}{RGB}{210,145,178}
\definecolor{myvioletc}{RGB}{45,130,60}
\definecolor{mygreen}{RGB}{34,139,34}
\definecolor{myorange}{RGB}{255,165,0}
\definecolor{OliveGreen}{RGB}{85,107,47}
\definecolor{NavyBlue}{RGB}{0,0,128}
\usepackage{comment}
\usetikzlibrary{decorations.markings}
\usetikzlibrary{decorations.pathmorphing}
\usepackage{tikz-feynman}
\usepackage{standalone}

\usepackage{listings}
\usepackage{hyperref}          

\usepackage[accepted]{icml2025}

\usepackage{amsmath}
\usepackage{amssymb}
\usepackage{mathtools}
\usepackage{amsthm}
\usepackage{enumitem}

\usepackage[capitalize,noabbrev]{cleveref}

\theoremstyle{plain}

\theoremstyle{definition}

\theoremstyle{remark}

\usepackage[textsize=tiny]{todonotes}

\usepackage{listings}
\icmltitlerunning{Automatic Karyotyping: From Metaphase Image to Diagnostic Prediction}

\newcommand{\pg}[1]{{\bf #1.}}

\begin{document}

\twocolumn[
\icmltitle{Automatic Karyotyping: From Metaphase Image to Diagnostic Prediction}





\icmlsetsymbol{equal}{*}
\icmlsetsymbol{equal_senior}{\dag}

\begin{icmlauthorlist}
\icmlauthor{Zahra Shamsi}{equal,gr}
\icmlauthor{Isaac Reid}{equal,gr,cam}
\icmlauthor{Drew Bryant}{equal,gr}
\icmlauthor{Jacob Wilson}{equal,fh}
\icmlauthor{Xiaoyu Qu}{fh}
\icmlauthor{Avinava Dubey}{gr}
\icmlauthor{Konik Kothari}{gr}
\icmlauthor{Mostafa Dehghani}{gr}
\icmlauthor{Mariya Chavarha}{gr}
\icmlauthor{Valerii Likhosherstov}{gr}
\icmlauthor{Brian Williams}{gr}
\icmlauthor{Michael Frumkin}{gr}
\icmlauthor{Anurag Arnab}{gr}
\icmlauthor{Adrian Weller}{cam,alan}
\icmlauthor{Frederick Appelbaum}{fh,wash}
\newline
\icmlauthor{Krzysztof Choromanski}{equal_senior,gdm}
\icmlauthor{Ali Bashir}{equal_senior,gr}
\icmlauthor{Min Fang}{equal_senior,fh,wash}
\end{icmlauthorlist}

\icmlaffiliation{gr}{Google Research, USA}
\icmlaffiliation{cam}{University of Cambridge, UK}
\icmlaffiliation{fh}{Fred Hutchison Cancer Center, Seattle, WA, USA}
\icmlaffiliation{alan}{Alan Turing Institute, UK}
\icmlaffiliation{wash}{University of Washington, Seattle, WA, USA}
\icmlaffiliation{gdm}{Google DeepMind, USA}

\icmlcorrespondingauthor{Krzysztof Choromanski}{kchoro@google.com}
\icmlcorrespondingauthor{Min Fang}{mfang@fredhutch.org}

\icmlkeywords{Machine Learning, ICML}

\vskip 0.2in
]



\printAffiliationsAndNotice{\icmlEqualContribution} 

\begin{abstract}
We present a machine learning method capable of accurately detecting chromosome abnormalities that cause blood cancers directly from microscope images of the metaphase stage of cell division.
The pipeline is built on a series of finetuned Vision Transformers.
Current state of the art (and standard clinical practice) requires expensive, manual expert analysis, whereas our pipeline takes only 15 seconds per metaphase image.
Using a novel pretraining-finetuning strategy to mitigate the challenge of data scarcity, we achieve a high precision-recall score of \textbf{94\% AUC} for the clinically significant del(5q) and t(9;22) anomalies.
Our method also unlocks zero-shot detection of rare aberrations based on model latent embeddings.
The ability to quickly, accurately, and scalably diagnose genetic abnormalities directly from metaphase images could transform karyotyping practice and improve patient outcomes. 
We will make code publicly available.\vspace{-3mm}
\end{abstract}

\vspace{-5mm}
\section{Introduction}
\label{sec:intro}
Chromosome analysis is essential for the diagnosis of genetic disorders and cancer.
\emph{Karyotyping} -- where clinicians arrest, stain and image chromosomes undergoing the metaphase stage of the cell cycle and assess them for genomic abnormalities -- is the first-line test for haemotological malignancies (blood cancers) like leukemia and lymphoma \citep{nguyen2022complex, mrozek2008cytogenetic}.
Despite playing a crucial role in driving therapeutic decisions, preparing and analysing karyograms (karyotype images) is a time-consuming, manual process, taking a skilled technologist many hours of hands-on analysis work per specimen.
Efforts to automate karyotyping have historically fallen short.

Two key technical barriers have frustrated previous efforts to automate chromosome aberration detection. 
\begin{enumerate}[leftmargin=*, itemsep=0pt, topsep=1pt]
    \item \emph{Data scarcity.} High-quality, labelled biomedical data is in general very scarce. 
    This problem is particularly acute for karyotyping because chromosome abnormalities with haematological malignancies can be rare, complex and diverse \citep{akkari2022guiding}. 
    The challenge is compounded by privacy concerns: patient data from cancer treatments is often considered very sensitive \citep{price2019privacy}.
    Whilst finetuning is the usual recourse for this small-data regime, it is not obvious how to define a suitable pretraining task to initialise the model weights.
    \item \emph{Image preprocessing.}
    Even if one is able to accurately predict whether an \emph{individual} chromosome is aberrant using an anomaly detection model, one still needs to detect and segment it from a bigger image of the cell undergoing metaphase in the first place. 
    Metaphase images are often very messy, with occlusions and spurious debris, so this is a technical challenge in its own right \citep{munot2011automated, remya2019preprocessing}.
\end{enumerate}
Previous progress towards automatic aberration detection has been limited.
Sec.~\ref{sec:related_work} provides full details.

\pg{Our contributions}
In this paper, we present (to our knowledge) \emph{the first end-to-end machine learning model for automatic karyotyping}, capable of taking a raw image of a cell undergoing metaphase and returning diagnostic predictions for chromosomal abnormalities such as del(5q) and t(9;22).
Using Vision Transformers \citep[ViTs;][]{dosovitskiy2020image}, it achieves near expert-level performance with \textbf{94\% AUC} for the precision-recall curve for both anomalies.
Moreover, our model takes only \textbf{15 seconds} for inference per metaphase image -- orders of magnitude less than current clinical practice which requires expensive human expertise.

\begin{figure*}[ht]
    \centering
    \includestandalone{pipeline_schematic} \vspace{-5mm}
    \caption{\textbf{Overall schematic.} The model takes a microscope image of a stained cell undergoing metaphase as an input. 
    OWL-ViT \citep{minderer2022simple}, finetuned on chromosome data, is used to predict bounding boxes.
    SAM \citep{kirillov2023segment} used to segment each chromosome therein.
    Each segmented chromosome is fed into a chromosome identity (`chrmID') ViT which predicts its number (1-22, X or Y).
    Candidate chromosome 5s are fed into a ViT that detects del(5q), and candidate chromosome 9 and 22s are fed into networks that detect t(9;22).
    This is straightforwardly extended to incorporate further structural aberrations.
    Multiple metaphase image predictions are aggregated to predict each patient's karyotype.
    }
    \label{fig:overall schematic}
\end{figure*}

To overcome the first challenge of scarce aberration data, we introduce a novel pretraining task based on predicting chromosome identity (1-22, X or Y). 
This is found to be essential for good diagnostic performance.
It may be of independent interest for machine learning in cytogenetics.
For the second challenge of complex image preprocessing, we finetune OWL-ViT \citep{minderer2022simple} on publicly available metaphase images with chromosome bounding boxes, and use Segment Anything \citep{kirillov2023segment} to extract the individual chromosomes from these boxes.
Notably, our Transformer-based model remains accurate even in few-shot learning scenarios with extremely rare aberrations.
We show that it also unlocks zero-shot identification of abnormal chromosomes based on their latent embeddings -- a powerful new diagnostic tool.
To our knowledge, this is the \textbf{first paper} to apply ViTs to chromosome aberration detection. 
We will make all code publicly available.

\pg{Impact} Clinicians report a positive qualitative experience using our model. 
Slow, manual processes have historically bottlenecked karyotyping to $\sim20$ metaphase images per patient.
The ability to scale beyond this with an accurate, fast machine learning tool could have a transformative impact on the treatment of haematological malignancies, providing faster test turnaround times and more detailed clinical understanding of the clonal architecture of the disease.  

\pg{Remainder of paper} 
Sec.~\ref{sec:primer} provides a short primer on cytogenetics and karyotyping, and discusses related prior work. 
Sec.~\ref{sec:automatic_karyotyping} gives an overview of our model pipeline. 
Sec.~\ref{sec:experiments} provides detailed experimental evaluations, demonstrating the ability of our model to quickly and accurately diagnose haematological malignancies directly from metaphase.

\vspace{-3mm}
\section{A brief primer on karyotyping} \label{sec:primer}
\emph{Chromosomes} are threadlike structures of nucleic acids and protein found in the nucleus of cells, carrying genetic information.
Healthy cells have 46 in total: 44 autosomes arranged in 22 homologous pairs, and two sex chromosomes (XX or XY). 
During \emph{metaphase}, a stage of cell division, they condense and move together, aligning in the centre of the cell.
Chromosomes are referred to as \emph{abnormal} if their structure is modified by aberrations like translocations, inversions, deletions, insertions, duplications and amplifications.
Cells may also exhibit numerical changes -- gains or losses of entire chromosomes -- which is called \emph{aneuploidy}.
These abnormalities drive and are used as diagnostic markers of disease, especially haemotological malignancies (blood cancers) like leukemia and lymphoma.

\emph{Karyotyping} is the process by which cytogeneticists arrest, stain and image chromosomes \citep{ozkan2020genetics}.
They count them and analyse their \emph{banding} (pattern of light and dark strips due to staining) for signs of aberrations. 
For instance, \emph{del(5q)}, deletion of the long arm of chromosome 5, is associated with diseases such as myelodysplastic syndromes and other myeloid neoplasms \citep{gurnari2022myelodysplastic}.
Meanwhile, \emph{t(9;22)}, the translocation of genetic material betweeen the long arms of chromosomes 9 and 22, is the diagnostic hallmark of chronic myelogenous leukemia. It is also associated with acute lymphoblastic leukemia, and more rarely with other acute leukemias \citep{tasian2017philadelphia}. 

Karyotype analysis for neoplasm workup (cancer testing) typically requires cytogeneticists to analyse 20 cells per patient case. 
This is performed either directly on metaphase images, or on \emph{karyograms}. 
The karyogram is produced by segmenting chromosomes out of the metaphase image and arranging them based on their identity -- a manual, laborious process. 
Depending on the complexity of the case, karyograms may be needed for up to all 20 metaphases. 
Based on the analysis of the chromosomes, the patient case is then assigned a \emph{karyotype}: an alphanumeric code indicating the diagnosis. 
For example, `\lstinline{46,XX,del(5q)}' denotes a patient with the full chromosome count, of which two are X. 
The long arm of one of the chromosome 5s has been deleted. 
Typically, an abnormality is included in the karyotype if it is present in at least two of the metaphase cells.
In this case, it is referred to as \emph{clonal} \citep{rack2019european}.

\vspace{-2mm}
\subsection{Related work} \label{sec:related_work}
Despite garnering interest from as early as the 1980s \citep{zeidler1988automated}, the technical challenges described in Sec.~\ref{sec:intro} have made progress towards automated karyotyping slow.
Early methods tackled segmentation using classical computer vision algorithms \citep{ji1989intelligent, ji1994fully, sugapriyaa2018segmentation, altinsoy2020fully}.
Using hand-crafted feature extractors, they achieved modest success for well-separated chromosomes but struggled with occlusions and adhesions. 
Recent work has replaced these tools with convolutional neural networks (CNNs) \citep{saleh2019overlapping,bai2020chromosome} -- especially \emph{You Only Look Once} \citep[YOLO;][]{redmon2016you} which performs well even on challenging examples \citep{tseng2023open}.

In parallel, some progress has been made towards automated chromosome classification \citep{abid2018survey,andrade2020study}. 
Using CNNs on pre-segmented images of individual chromosomes, one can accurately predict each chromosome's identity (1-22, X or Y) \citep{yang2023chromosome}.
This can be extended to differentiate some common structural abnormalities.
However, demanding such extensive image preprocessing -- rather than working directly with metaphase images -- limits such tools' practical utility.

Preliminary attempts have been made to combine chromosome segmentation and classification end-to-end, including for diagnosing diseases due to extra or missing chromosomes \citep{wang2024fully}.
However, to our knowledge, no previous end-to-end system is capable of diagnosing structural chromosomal abnormalities like del(5q) and t(9;22).
Likewise, \emph{de novo} detection methods, designed for rare aberrations with little or no training data, are lacking.
Evaluations are often contrived, benchmarking on small datasets of `easy' examples \citep{altinsoy2019raw}.
There is a stark absence of practical, robust, end-to-end tools, leveraging modern Transformer architectures and pretraining-finetuning strategies. 
Our paper seeks to remedy this.

\vspace{-3mm}
\section{Automatic karyotyping: an overview} \label{sec:automatic_karyotyping}
In this section, we describe our novel state of the art model for predicting patient karyotypes directly from images of cells undergoing metaphase cell division.
It consists of two components: 1) \emph{image preprocessing}, which detects and segments each individual chromosome from the cell, and 2) \emph{chromosome identity and aberration prediction}, which detects disease amongst these segmented instances.
The chromosome number and aberration predictions are combined to make a prediction for the cell, then predictions for multiple ($\sim 20$) cells are aggregated to predict the patient's overall karyotype.
Fig.~\ref{fig:overall schematic} summarises this in a schematic.
The remainder of Sec.~\ref{sec:automatic_karyotyping} provides an overview of our techniques and motivation; full experimental evaluations are in Sec.~\ref{sec:experiments}

\subsection{Image preprocessing: OWL-ViT and SAM}
To detect and segment individual chromosomes, we use a pair of Vision Transformers (ViTs).
This task has historically been tackled using classical computer vision algorithms, or more recently convolutional neural networks (CNNs).
To our knowledge, this is the \textbf{first application} of ViTs to this problem.

\pg{Chromosome detection}
Our first task is to detect the chromosomes within the microscope image of a metaphase cell.
To achieve this, we use OWL-ViT \citep{minderer2022simple}: an open-vocabulary object detection network based on a Vision Transformer (ViT) architecture, pretrained at scale using image-text contrastive loss then finetuned to predict bounding boxes for text queries.

For the text query, we input the word `\lstinline{chromosomes}'.
The performance of the base OWL-ViT model is very poor because this specialised vocabulary is rare in the training data; the model typically fails to detect anything.
To resolve this, we fine-tune OWL-ViT using $5000$ publicly available metaphase images, annotated with bounding boxes \citep{tseng2023open}.
In particular, we train for an extra $1000$ steps, taking a linear cooldown with initial learning rate $2 \times 10^{-5}$.
This drastically improves performance.

\pg{Chromosome segmentation}
Having obtained predictive bounding boxes, our next step is to remove overlapping sections of other chromosomes and spurious cell debris. 
To achieve this, we use Meta's Segment Anything Model \citep[SAM;][]{kirillov2023segment}, taking the coordinates of each bounding box as a prompt.
This effectively isolates the pixels of each individual chromosome, ready to be fed into the anomaly detection trunk.
In contrast to OWL-ViT, we find that SAM does not require any special finetuning for our setting.

\subsection{Chromosome identification and anomaly prediction} \label{sec:chromosome_id_and_anom_pred}
Having obtained 46 segmented chromosome instances (or more/fewer for patients with aneuploidy), the next task is to predict whether each one is healthy or anomalous.
This is challenging because anomalies can be complex, diverse and rare.
We achieve this task using another ViT, requiring a \textbf{novel dataset} and a \textbf{novel pretraining} strategy.

\pg{Dataset curation} 
The training dataset was assembled using results collected from five years (2016-2020) of clinical testing performed by the 
Fred Hutchinson Cancer Center (FHCC) Cytogenetics Laboratory. 
As part of routine testing, metaphase images captured by an automated imaging system were processed (cropped, segmented and rearranged) into karyograms by experienced cytogenetic technologists -- a process typically taking approximately 2 hours to days per sample, depending on the complexity level. 
Of the $42049$ karyograms collected from $9207$ bone marrow and peripheral blood specimens by $4711$ patients with hemotological malignancies, approximately $70\%$ were classified as normal and $30\%$ as abnormal. 
Each karyogram was labelled with the corresponding patient diagnosis and an alphanumeric karyotype, following the standard International System for Human Cytogenetic Nomenclature (ISCN).
From every karyogram, each constituent chromosome image was then isolated, centered on a white background, and cropped to $199 \times 99$ pixels.
Finally, every such chromosome image was labelled by its identity (1-22, X or Y), normal/abnormal, and where present the specific abnormality -- data that can be inferred from the chromosome's position in the karyogram and the karyotype/expert assessment.
An independent follow-up dataset from two additional years (2021-2022) of clinical testing, including 3736 specimens from 1608 unique patients, was also assembled to evaluate model predictions. 
This was approved by the Institutional Review Board of the 
Fred Hutchison Cancer Center.

\begin{figure*}[ht]
    \centering
    \includegraphics[width=0.95\textwidth]{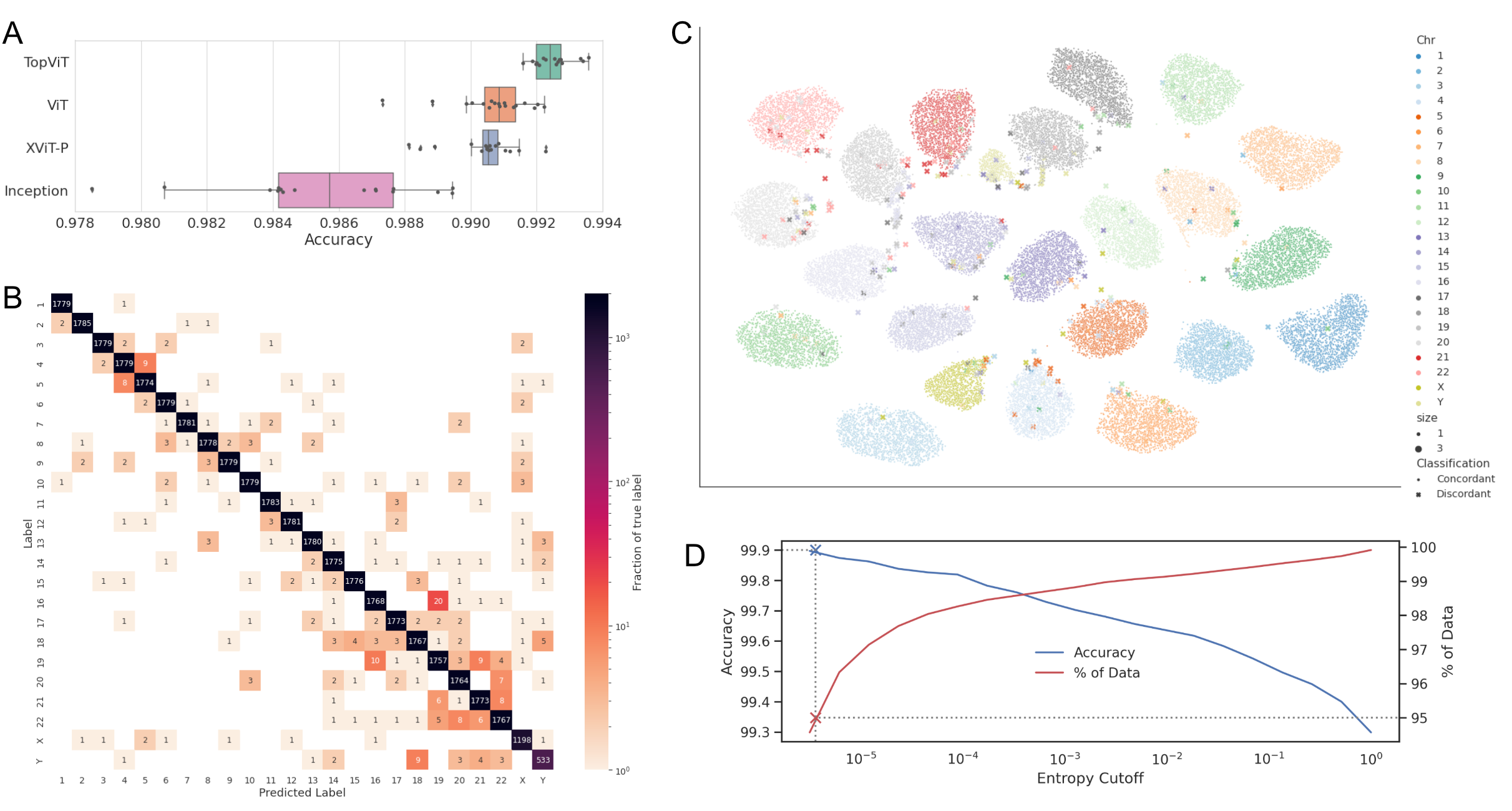} \vspace{-5mm}
    \caption{\textbf{Chromosome identity prediction}.
    \emph{(A)} Accuracy of different models for chromosome identity prediction (1-22, X or Y), from pre-segmented, aligned and cropped chromosome images taken from respective karyograms.
    Performer with block-Toeplitz masking (`TopViT') performs best, followed by softmax attention (`ViT') then regular Performer (`XViT-P'). 
    The CNN (Inception) is consistently worst.
    Black dots are hyperparameter instantiations. 
    \emph{(B)} Confusion matrix for the best chromosome identification model on a test set. 
    \emph{(C)} UMAP projection of the last intermediate layer (pre-logits) for the best-performing ViT model on a test set. 
    Each point is coloured by its ground truth label, with predictions that disagree showing their label enlarged ten-fold and marked `$\times$'. 
    \emph{(D)} Classification accuracy (left $y$-axis) and percentage of remaining data (right $y$-axis) after removing high-entropy predictions, as a function of entropy cutoff ($x$-axis). 
    For example, removing predictions with entropy $3.6 \times 10^{-6}$ or greater increases the accuracy to 99.9\% by removing 5\% of the data.}
    \label{fig:chrmid-fig}
\end{figure*}

\pg{Chromosome identification}
To predict each chromosome's identity (1-22, X or Y), we train a ViT on our curated 2016-2020 dataset.
We use the healthy chromosomes, which contain roughly equal counts of each number (apart from X and Y).
For the attention mechanism, we consider three variants:
\begin{enumerate}[leftmargin=*, itemsep=-1pt, topsep=0pt]
\item \emph{Softmax.} Regular, full-rank softmax attention. 
Takes $\mathcal{O}(N^2)$ time, with $N$ the number of tokens (patches) \citep{vaswani2017attention,dosovitskiy2020image}.
\item \emph{Performer.} Low-rank attention.
Takes $\mathcal{O}(N)$ time, typically at the cost of worse performance \citep{choromanski2020rethinking}.
\item \emph{Performer with block-Toeplitz masking.} Low-rank attention, but enhanced with an efficient relative position encoding (RPE) strategy (`topological masking') to incorporate structural inductive bias and boost performance.
Takes $\mathcal{O}(N \log N)$ time \citep{choromanski2022block}.
\end{enumerate}
Whilst softmax has become the standard choice in Transformers, we also include the subquadratic Performer mechanisms because for us \emph{efficiency} is also important.
The chromosome identification model will typically be applied 46 times per metaphase image and thus at least $920$ times per patient, so high-throughput is  desirable.
Moreover, our model is intended to be deployed in settings which may not have access to specialised accelerator hardware.
Empirically, the third variant, which augments regular linear attention with an improved RPE mechanism, strikes the right balance between performance and inference time.
Sec.~\ref{sec:anomaly_detection_network} describes our results in detail; for now, we note that all models achieve very high classification accuracy ($>99\%$) and substantially outperform a CNN baseline \citep[InceptionV3;][]{szegedy2016rethinking,vajen2022classification}.

\pg{Anomaly detection}
The ability to detect, segment and classify chromosomes based on their number already unlocks automated diagnosis of \emph{aneuploidy}, where cells have additional or missing chromosomes.
This is a common cause of genetic disorders and a marker of cancer; about 68\% of solid tumours are aneuploid \citep{duijf2013cancer}.
However, for truly automatic karyotyping, we would also like to directly detect specific structural chromosomal abnormalities like deletions and translocations, namely del(5q) and t(9;22), from the metaphase images.

This is more challenging than chromosome identification because labelled data for aberrant chromosomes is even more scarce than for healthy chromosomes. 
Amongst our curated dataset, we have only $1247$ labelled instances of del(5q) chromosomes and $1094$ instances of t(9;22) chromosomes.
Other abnormalities are even rarer, with $<50$ examples respectively.
This is not enough to train a ViT from scratch, risking overfitting and failing to capture the diverse nature of chromosomal abnormalities.

To mitigate this, we propose a \emph{novel pretraining strategy}.
We initialise the anomaly detection ViT's weights using the chromosome identification model, then warm-start and resume training for a further 1000 epochs.
We use a tiny learning rate to prevent overfitting.
Training separate anomaly ViTs for del(5q) and t(9;22), the latter for both chromosome 9 and chromosome 22, we obtain very high precision and recall, consistently surpassing that achieved by training from scratch. 
Full details are in Sec.~\ref{sec:from_metaphase} and App.~\ref{app:hypers}.

\pg{Broader significance} 
As a further remark: whilst this paper considers the specific problem of chromosome anomaly prediction, we anticipate that the approach we have adopted -- pretraining on a surrogate task with comparatively abundant, healthy examples, and then finetuning on a small, sensitive dataset -- may be of broader interest for machine learning applications in medicine.

\subsection{From metaphase to karyotype} \label{sec:get_karyotype_pred}
The final stage of the automatic karyotyping pipeline takes the output logits from the chromosome identity and anomaly networks for all $N$ metaphase images, and aggregates them to predict the overall patient karyotype.
There are many possible ways to achieve this.
For instance, one could train yet another neural network.
However, in the interests of interpretability, to avoid overfitting, and to better model clinical practice (where at least two aberrant cells are needed to define a clone \citep{mcgowan2020iscn}), we simply \emph{count} the number of predicted anomalous metaphase images per patient -- i.e.~the number of cells for which at least one of the candidate chromosome 5s is predicted to be del(5q), and likewise for t(9;22). 
This design choice works very well in practice (see Sec.~\ref{sec:from_metaphase}).
Investigating the merits and drawbacks of more sophisticated strategies is left to future work.

\section{Experiments} \label{sec:experiments}

Sec.~\ref{sec:automatic_karyotyping} detailed the components of our Transformer-based model for automatic karyotyping, from chromosome detection and segmentation to identity and anomaly prediction, followed by final karyotype classification. 
In this section we empirically investigate this pipeline, comparing to baselines, performing ablations over model and training parameters, and demonstrating the ability of our approach to accurately and quickly diagnose haematological malignancies.

\begin{figure*}[ht]
    \centering
    \includegraphics[width=0.95\textwidth]{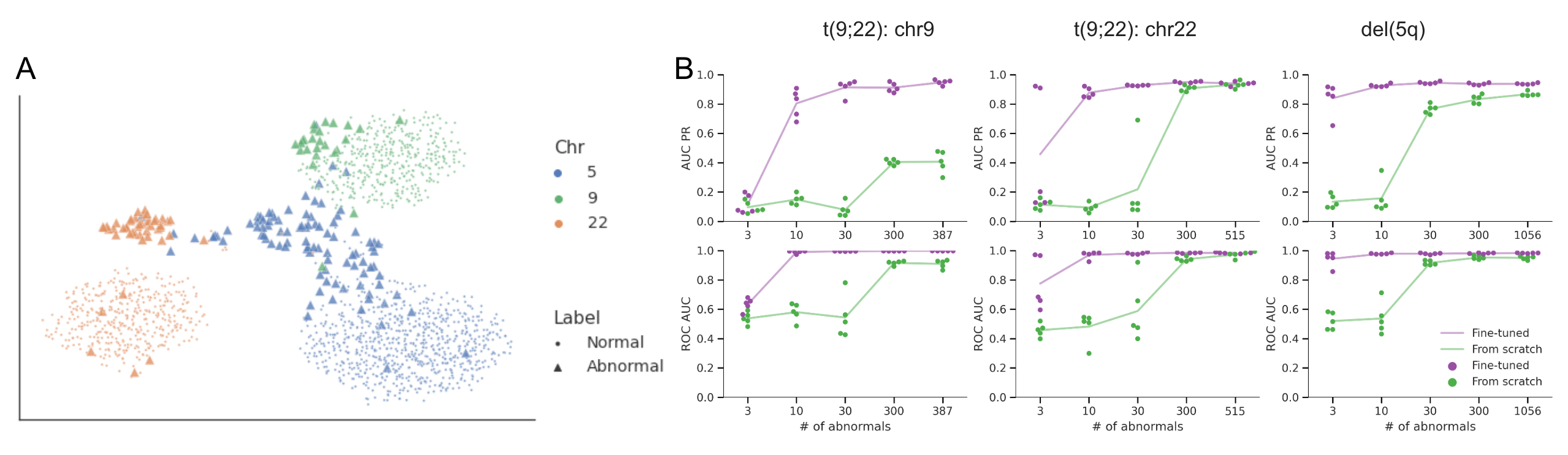} \vspace{-5mm}
    \caption{\textbf{Aberration detection for del(5q) and t(9;22) anomalies}.
    \emph{(A)} UMAP projection of normal and abnormal chromosomes 5, 9 and 22, constructed using the previous chromosome identification network from Sec.~\ref{sec:chrmid_model}.
    Healthy and anomalous examples are already fairly well-separated, motivating initialising the anomaly detection ViT with these weights or a \emph{de novo} strategy. 
    \emph{(B)} Precision-recall area under curve (AUC-PR) and reciever-operator characteristic area under curve (ROC-AUC) -- two measures of diagnostic test accuracy -- plotted as a function of train dataset size for different anomalies.
    Green markers show performance when trained from scratch, whereas purple markers show performance when finetuned. 
    The latter is consistently much better and substantially reduces the amount data needed for good performance.
    }
    \label{fig:abb_pred}
\end{figure*}

\subsection{Pretraining: predicting chromosome identity} \label{sec:chrmid_model}
Here, we evaluate model performance for chromosome identification in karyotype images. 
Panel A of Fig.~\ref{fig:chrmid-fig} shows the average classification accuracy on a held out test set.
For each of the ViT attention mechanisms described in Sec.~\ref{sec:chromosome_id_and_anom_pred} we perform hyperparameter sweeps over the learning rate, optimiser, batch size, patch size, number of self-attention heads, size of model hidden state, multilayer perceptron (MLP) dimensionality, number of layers, and (for the Performer variants) kernel type and number of kernel features. 
We take a compound learning rate with linear warmup, a constant base rate, then decay. 
Exhaustive details are provided in App.~\ref{app:hypers}.
For the CNN baseline, we train an off-the-shelf InceptionV3 model \citep{vajen2022classification,szegedy2016rethinking}. 
Each black dot represents a specific instantiation of the hyperparameters.
We find that the Performer with block-Toeplitz masking generally works best, halving the number of classification errors compared to the CNN baseline.

\pg{Confusion matrix}
Panel B of Fig.~\ref{fig:chrmid-fig} shows the \emph{confusion matrix} -- the distribution of errors between chromosome labels -- for the best-performing Performer plus topological masking (`TopViT') model. 
The most discordant predictions are between chromosomes of similar sizes. 
They are balanced, meaning the plot is roughly symmetric about its diagonal axis.
Some of the most frequent mistakes are between chromosomes that appear visually similar to humans, e.g.~swapping $4$ and $5$ or $21$ and $22$.
Interestingly, others appear to be specific to the model, e.g. $16$ and $19$ or $20$ and $22$.

\pg{UMAP embeddings}
To better understand confusion matrix, in panel C of Fig.~\ref{fig:chrmid-fig} we show a \emph{Uniform Manifold Approximation and Projection} \citep[UMAP;][]{mcinnes2018umap} -- a dimensionality reduction technique to visualise high-dimensional data -- for the last layer of the chromosome identification Transformer model. 
Incorrect predictions are tightly localised, suggesting the same underlying mechanism for systematic swaps. 
Moreover, chromosome clusters are generally well-separated, with discordancies mostly located at the peripheries. 
This suggests that mismatched predictions are considered to be less `canonical' examples for the predicted chromosomes.

\pg{Entropy and data cleaning}
To push this further, we compute the entropy of the 24-class probability vector for all model predictions. 
Discordant predictions generally have much higher entropy than concordant predictions, indicating greater model uncertainty. 
We introduce a simple \emph{entropy-based filter}, discarding examples for which $-\sum p \log p$ exceeds some prescribed threshold.
For instance, setting the cutoff at $3.6 \times 10^{-6}$, the accuracy reaches $99.9\%$ by dropping $5\%$ of the data. 
Panel D of Fig.~\ref{fig:chrmid-fig} shows model performance and percentage of remaining data for different choices of cutoff.
Performing an expert, manual re-examination of all discordant predictions (both low and high entropy), we find that most (72\%) of the low-entropy discordancies are due to mislabelled training data.
Meanwhile, most (90\%) of the high-entropy discordancies are due to low-quality data (occluded, folded, or bent chromosomes, with poor banding, poor morphology or cellular debris). 
Correcting mislabels and dropping low-quality images, we achieve a remarkably high revised accuracy of $\mathbf{99.5\%}$.

\subsection{Finetuning: detecting chromosome aberrations} \label{sec:anomaly_detection_network}
Next, we evaluate the performance of our model for aberration detection in karyograms.
To begin we consider \textbf{del(5q)}, deletion of the long arm of chromosome 5, and \textbf{t(9;22)}, translocation between the long arms of chromosomes 9 and 22.
These are the most abundant abnormalities in our curated dataset and are of substantial medical interest.

\pg{UMAP embeddings} Panel A of Fig.~\ref{fig:abb_pred} shows the UMAP embeddings of the final layer of the chromosome identification network from Sec.~\ref{sec:chrmid_model}, for healthy and anomalous chromosomes 5, 9 and 22. 
Even without any specific extra training, aberrant examples already appear well-separated, motivating using finetuning.

\begin{figure*}[ht]
    \centering
    \includegraphics[width=0.9\linewidth]{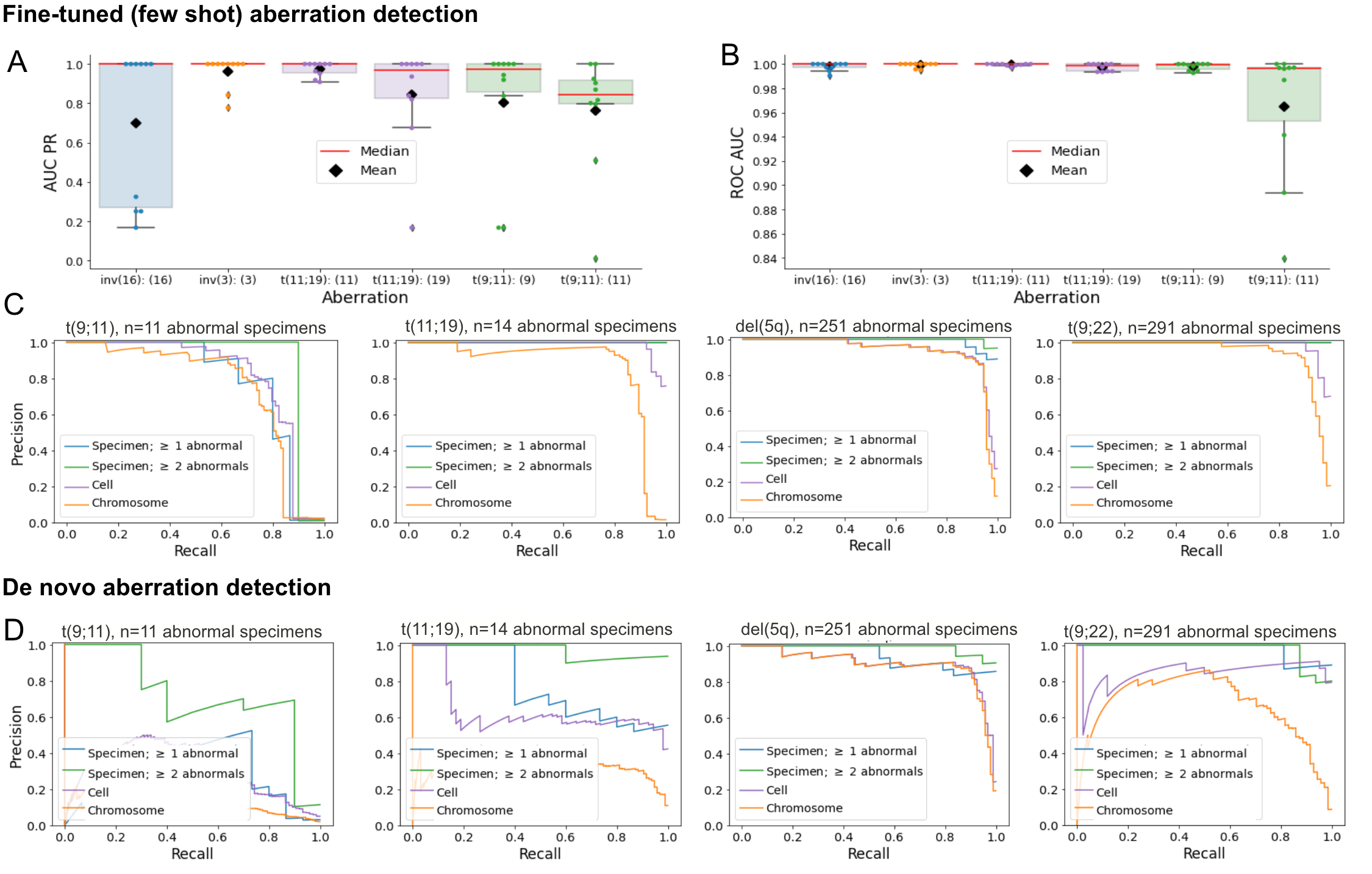} 
    \caption{\textbf{Model performance of rare aberrations and precision-recall curves when aggregating predictions across cells and specimens.}
    10-fold cross-validation performance for \emph{(A)} AUC PR and \emph{(B)} ROC AUC. 
    Each boxplot corresponds to a distinct cross-validation set for each chromosome involved in an aberration, and is coloured by aberration. 
    Individual points for folds and averages (black diamonds) are overlaid on each boxplot. 
    \emph{(C)} Precision-recall curves for t(9;11), t(11;19), del(5q), and t(9;22), at the individual chromosome image level (orange) or aggregated at the cell (purple) or specimen levels. 
    For specimen level $\geq$ 1 abnormal (blue) the single highest probability abnormal chromosome was used, for specimen level $\geq$ 2 abnormals (green) the second highest probability abnormal chromosome was used. 
    Similarly, \emph{(D)} shows precision-recall for \emph{de novo} aberration detection based on distance to $N$-nearest chromosomes (here 50th) for t(9;11), t(11;19), del(5q), and t(9;22), respectively.
    }\label{fig:rare_anomalies_result} \vspace{-4mm}
\end{figure*}

\pg{Finetuning for accurate anomaly detection}
Meanwhile, panel B of Fig.~\ref{fig:abb_pred} shows two popular measures of diagnostic test performance -- precision-recall area under curve (AUC-PR) and reciever-operator characteristic area under curve (ROC-AUC) -- plotted as a function of train dataset size for each anomaly.
To vary the train dataset size, we randomly subsample.
In every case, we evaluate on the same test set.
The purple line shows the results with a finetuned network, whereas the green line shows the results when trained from scratch with randomly initialised weights.
The finetuned models not only perform substantially better, but also \textbf{maintain high AUCs even with just tens of examples}.
They are extremely data efficient.
This is in stark contrast to models trained from scratch, which need hundreds of training examples to achieve good predictive performance.
This demonstrates the crucial practical importance of our new pretraining-finetuning strategy.

\pg{Low-frequency anomalies}
Encouraged by the strong performance of our finetuned models for detecting del(5q) and t(9;22) with very little training data, we now consider other infrequent aberrations in our curated dataset: inv(16), inv(3), t(11;19) and t(9;11). 
These are critical for disease diagnosis and prognosis, and represent diverse rearrangements of genetic material, prevalence, and difficulty levels for microscopic assessment.
In each case we have $<50$ examples; see App.~\ref{app:anomalies_table_sec} for details.
Given the limited sample sizes, we compute the average performance across 10-fold cross 
validation. 
The upper half of Fig.~\ref{fig:rare_anomalies_result} shows the results, finetuning as described in the paragraphs directly above.
Individual performance is often lower than for the more frequent, easier del(5q) and t(9;22) cases, but ROC-AUC is still found to be high across all rare aberrations and median and mean AUC-PR exceed 0.8 and 0.7 respectively.
In panel C, we also consider aggregating predictions within cells (purple) and across cells (blue and green) from a patient specimen, classifying as positive if at least one example (or two examples for `specimen $\geq 2$') is predicted to be anomalous.
Naturally, this improves performance.


\pg{Identifying aberrant chromosomes \emph{de novo}}
Lastly, given the suggestive UMAP results in Fig.~\ref{fig:abb_pred} without any finetuning, we attempt to identify aberrant chromosomes zero-shot. 
Using the chromosome identification ViT from Sec.~\ref{sec:chrmid_model}, we take the latent embedding vectors for each test chromosome and all normal chromosomes from training.
We calculate the Euclidean distance between each test and all training chromosomes, then select the distance to the nearest neighbour. 
Using the set of distances, we predict aberrations. 
Generating PR curves, the performance is comparable to finetuning across all aberrations; see Fig.~\ref{fig:rare_anomalies_result}D.
Remarkably, the latent representations from the final layer of the  ViT used to predict chromosome identity can already be used to distinguish whether an example is `canonical' or aberrant, permitting zero-shot detection of rare structural anomalies.

\vspace{-2mm}
\subsection{Putting it together: automatic karyotyping} \label{sec:from_metaphase}
Equipped with modules capable of detecting, segmenting and classifying chromosomes, our final task is to assemble the full pipeline and make predictions for patient karyotypes, as per Fig.~\ref{fig:overall schematic}. 
Given the strong results earlier in Sec.~\ref{sec:anomaly_detection_network}, we focus on the del(5q) and t(9;22) anomalies.

\pg{Data augmentation} In practice, we find that making predictions for the outputs of automatic detection and segmentation with OWL-ViT and SAM is more difficult than the preprocessed examples in our curated dataset, which have been cropped from a careful, manually-assembled karyogram. 
As such, we retrain the chromosome identification and anomaly detection networks with additional data augmentation (image rotation, resizing, cropping, and adjustment of contrast, brightness and sharpness) to ensure we generalise well to chromosomes automatically segmented from metaphase.
We also choose regular ViT attention for consistency across the pipeline.

\begin{figure}
    \centering
    \includegraphics[width=0.9\linewidth]{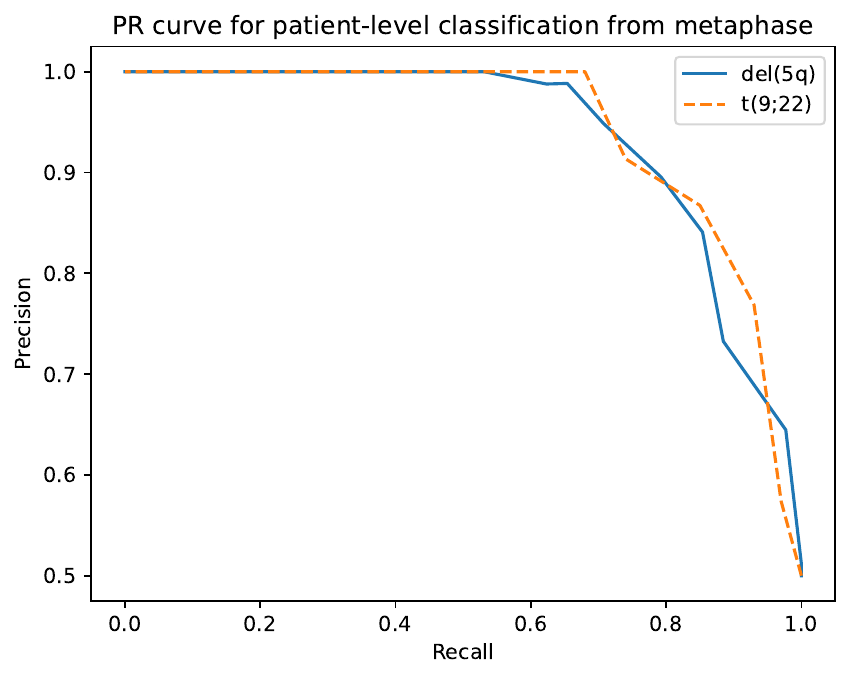} \vspace{-1mm}
    \caption{\textbf{Patient-level PR curve.}
    Precision-recall curve for patient-level detection of the del(5q) and t(9;22) chromosomal abnormalities, directly from $20$ microscope images of cells undergoing the metaphase stage of cell division. \vspace{-5mm}}
    \label{fig:patient-pr}
\end{figure}

\pg{Results} Fig.\ref{fig:patient-pr} shows the precision-recall curves, swept out by varying the threshold proportion of predicted anomalous metaphase images needed to classify a patient's karyotype as aberrant from $0$ to $1$ (see Sec.~\ref{sec:get_karyotype_pred}). We use a test set of $260$ del(5q) patients and $200$ t(9;22) patients, with an equal split between healthy and unhealthy examples.
This corresponds to $5200$ and $4000$ metaphase images respectively -- over $4\times10^5$ chromosomes in total. 
Table \ref{tab:karyotype-pr-aucs} reports the PR-AUCs for these anomalies, a key result of this paper.
\begin{tcolorbox}[colback=gray!10!white,colframe=gray!50!black,arc=0mm,boxrule=0.5pt]
\vspace{-5mm}
\begin{table}[H]
    \centering
    \small
    \begin{tabular}{c | c  c } 
    \toprule
    &  \textbf{del(5q)} & \textbf{t(9;22)} \\ \midrule
    PR-AUC & 0.938 & 0.944 \\
    \bottomrule
    \end{tabular}
    \caption{\textbf{Key result}. Final patient karyotype classification accuracies, given as areas under the precision-recall curves (Fig.~\ref{fig:patient-pr}).
   Uniquely, we classify \emph{directly} from microscope images of cells undergoing metaphase. 
    \vspace{-8mm}
    }
    \label{tab:karyotype-pr-aucs}
\end{table} 
\vspace{3mm}
\end{tcolorbox}

The PR-AUCs are remarkably high given that the pipeline is end-to-end, taking complex, possibly occluded metaphase images as inputs and outputting a patient diagnosis.
This is much more challenging than the experiments in Sec.~\ref{sec:anomaly_detection_network}, where the images of individual chromosomes have been manually, expertly prepared. 
Clinicians report that Table \ref{tab:karyotype-pr-aucs} approaches expert level performance.
We have achieved our core goal of accurately, reliably detecting anomalies responsible for haemotological malignancies directly from metaphase images.

\pg{Speed tests}
Running in Google Colab on a TPU, the entire del(5q)/t(9;22) automatic karyotyping pipeline outlined in Fig.~\ref{fig:overall schematic} takes $\textbf{15 seconds}$ per metaphase image. 
In contrast, an experienced technician might take $10$ minutes to crop and segment the chromosomes, prepare a karyogram, and analyse it to determine a diagnosis. 
This makes our automated method \textbf{faster by a factor of around 40} -- a very substantial time saving.
Reducing or removing expert human supervision will also make karyotyping cheaper and more accessible. 

The throughput of our pipeline is currently bottlenecked by the presence of multiple Transformer models.
Making inference even quicker, e.g.~by incorporating the segmentation step into OWL-ViT, is left as important future work.

\vspace{-2mm}
\section{Automatic karyotyping: a new paradigm}
Whilst the PR-AUC evaluations reported in Table \ref{tab:karyotype-pr-aucs} are impressive, perhaps even more important is our model's speed.
Fast, accurate predictions directly from metaphase images unlock karyotyping at unprecedented scale, considering far more specimens than possible with a tedious, manual process. 
Time and resource constraints have historically limited karyotyping to around 20 cells per patient, but the ability to consider many more samples could permit early screening of low-abundance subclonal lesions, unlikely to appear in a small sample size.
This could have a transformative impact on the treatment of cancer, unlocking earlier diagoses and providing clinicians with a better understanding of the clonal architecture amongst the patients' cells to inform the best treatment regimen.
Our model pipeline could also open up karyotyping where the access to expert clinicians is limited, improving the equitability of cancer treatment. 
We are excited to explore the full medical implications in future work.

\newpage
\section{Impact Statement}
This paper proposes a new algorithm for fast, accurate detection of chromosomal abnormalities that cause haemotological malignancies, directly from microscope images of dividing cells. 
We anticipate that it will improve patient outcomes and save doctors' time, providing a powerful new tool in the fight against cancer. 

\section{Acknowledgements and Contributions}

\pg{Acknowledgements}
We thank Mr Vijay Sureshkumar, former Director of Technology Strategy at the Fred Hutchison Cancer Center, for his tireless effort in making this collaboration with Google Accelerated Sciences possible. 
We also thank Mr. Andrew Carol, product lead at genomics team in Google AI, for resource coordination and helpful advice at various stages of the project.

\pg{Contributions}
MinF, AB, MichaelF, and DB conceptualised and designed the study.
JW, AB, DB, MC and ZS curated datasets. 
JW, XQ and MinF verified chromosome labeling. 
KK developed the Inception model. 
KC proposed a Transformer-based approach to the problem, applying Vision Transformers for the anomaly detection model. He developed the TopViT Transformer model, ran initial TopViT tests and detailed Performer tests.
ZS and DB developed and evaluated the Transformer-based models for chromosome identification and abnormality detection, with help from AD (pretraining/finetuning, model iteration), KC, MD, BW, VL. 
ZS, DB, AB, AD, KC analysed the results. 
ZS, DB, AB, AD, KC and JW prepared the visualizations for the report. 
All the aforementioned authors contributed to drafting, editing and approving an earlier draft the manuscript.
IR extended the pipeline to operate end-to-end on images of cells undergoing metaphase. 
This involved finetuning OWL-ViT, incorporating SAM, retraining the chromosome identification and anomaly detection models with data augmentation, and (with help from ZS) obtaining the experimental results in Sec.~\ref{sec:from_metaphase}. 
IR wrote the final version of the manuscript.

\bibliography{example_paper}
\bibliographystyle{icml2024}

\newpage
\appendix
\onecolumn
\section{Experimental details and extra results}
\subsection{ViT model and training details} \label{app:hypers}
\pg{Chromosome identification and aberration detection}
Table \ref{tab:vit_hypers} lists the hyperparameters swept over for the chromosome identification and aberration detection ViT models.
Table \ref{tab:final_vit} shows the final choices made for the experimental evaluations.
We build on the Scenic codebase.\footnote{Available at: \href{https://github.com/google-research/scenic}{\lstinline{https://github.com/google-research/scenic}}.}

\begin{table}[H]
    \centering
    \small
    \begin{tabular}{c | c  } 
    \toprule
    Parameter &  \{Values\}  \\ 
    \midrule
    Number of heads & 3,12,16 \\
    Number of hidden layers & 6,12,16,24 \\
    Hidden  layer size & 192, 768, 1024 \\
    MLP dimension & 384, 768, 1534, 3072 \\
    Patch size & $1 \times 1$, $2 \times 2$, $4 \times 4$, $8 \times 8$, $16 \times 16$ \\
    Batch size & 128, 256, 512, 1024, 2048, 4096 \\
    Optimiser & Adam, SGD \\
    \bottomrule
    \end{tabular}
    \caption{Hyperparameter sweeps for ViT models in Fig.~\ref{fig:chrmid-fig}A.
    The ratio of self-attention heads to hidden state size is fixed to $64$.
    }
    \label{tab:vit_hypers}
\end{table} 
\begin{table}[H]
    \centering
    \small
    \begin{tabular}{c | c  } 
    \toprule
    Parameter &  Value  \\ \midrule
    Number of heads & 12  \\
    Number of hidden layers & 16 \\
    Hidden layer size & 768 \\
    MLP dimension & 768 \\
    Patch size & $4 \times 4$ \\
    Batch size & 512 \\
    Pretrain steps (for aberration detection) & 359000 \\
    Loss & Cross Entropy \\
    Activation function & gelu \\
    Dropout rate & 0.1 \\
    Attention dropout rate & 0.1 \\
    Optimiser & Adam \\
    Compute resource & $8 \times 8$ TPUv3 \\
    Attention type & ReLU Performer + Block-Toeplitz  (Secs \ref{sec:chrmid_model} and \ref{sec:anomaly_detection_network})  \\
    & Softmax (Sec.~\ref{sec:from_metaphase}) \\
    \toprule 
    Learning rate & \\
    \midrule
    Schedule & Compound \\
    Factors & Constant $\times$ linear warmup $\times$ linear decay\\
    Base LR & $3 \times 10^{-3}$ (from scratch)\\
     & $4 \times 10^{-6} - 3 \times 10^{-5}$ (fine-tuning) \\
     Warm-up steps & 10000 (chromosome identification) \\ 
      & 1000 - 5000 (aberrations) \\
    End LR & $1 \times 10^{-6}$ \\
    Total steps & 359000 (chromosome identification) \\
     & 2000-6000 (aberrations) \\
    \bottomrule
    \end{tabular}
    \caption{ViT hyperparameters used in final chromosome identification and aberration detection models.
    Where a range of values is listed, we run an extra small hyperparameter sweep so the choice in Secs \ref{sec:chrmid_model} and \ref{sec:anomaly_detection_network} may differ to in Sec.~\ref{sec:from_metaphase}.
    }
    \label{tab:final_vit}
\end{table} 

\pg{Segment Anything Model}
For SAM \citep{kirillov2023segment}, we use a standard ViT-H vision encoder.
This has 32 layers with 16 heads, and latent representation dimensionality 1280 (i.e.~80 per head).
This totals about 636M parameters.

\pg{OWL-ViT}
For OWL-ViT \citep{minderer2022simple}, we again use an implementation from Scenic.\footnote{Available at: \href{https://github.com/google-research/scenic/tree/main/scenic/projects/owl_vit}{\lstinline{https://github.com/google-research/scenic/tree/main/scenic/projects/owl\_vit}}.}
We use the `v2' variant, using a standard ViT-L backbone with patch size 14. 
We finetune from a publicly available checkpoint available via the repository, training for 1000 steps with linear cooldown from an initial learning rate of $2 \times 10^{-5}$.

\subsection{Summary of chromosome aberrations evaluated} \label{app:anomalies_table_sec}
Table \ref{table:aberrations} shows the different chromosome aberrations present in our dataset, their respective prevalences, and various other properties of medical interest.
\begin{sidewaystable} 
\centering
\tiny
\begin{tabular}{ l l l l l l l l l l l }
\toprule
Aberration & Gene fusion or & Prevalence & \multicolumn{2}{c}{ \hspace{-5.5em}Clinical significance} & Difficulty & Aberration type\textsuperscript{1} & \multicolumn{2}{c}{Number of abnormal}   & \multicolumn{2}{c}{Numnber of normal} \\
 & rearrangement &  & Prognosis & Therapy &  level &  & specimens & chr. & specimens & chr. \\
\midrule
t(9;22)(q34;q11.2) & BCR::ABL1 & 100\% CML & Good & TKI & Easy & Inter-chromosomal & 160 & 1094 & 1242 & 9144 \\
& & 25\% Adult B-ALL & Poor & TKI &  &  &  &  &  &  \\
& & 2-4\% Pediatric B-ALL & Poor & TKI &  &  &  &  &  &  \\
& & $<1\%$ AML and MPAL & Poor & TKI suggested, along &  &  &  &  &  &  \\
& &  &  & with chemotherapy &  &  &   &  &  &  \\
\midrule
del(5q) & NA & 100\% MDS w/ isolated & Good & Lenalidomide & Easy to moderate & Intra-chromosomal & 251 & 1247 & 1482 & 7613 \\
 &  &  del(5q)&  &  & & unbalanced &  &  &  &  \\
 &  & 10-40\% Myeloid neoplasm & Variable\textsuperscript{2} & Variable & &  &  &  &  &  \\
\midrule
t(9;11)(p21.3;q23.3) & KMT2A::MLLT3 & 9-12\% Pediatric AML& Intermediate & Chemotherapy; HSCT & Moderate & Inter-chromosomal & 11 & 39 & 1092 & 3513 \\
 & & 2\% Adult AML &  &  &  &  &  &  &  &  \\
 & & $<1\%$ Other myeloid neoplasm &  & &  &  &  &  &  &  \\
\midrule
inv(16)\textsuperscript{3} & CBFB::MYH11 & 5-8\% AML & Good & Chemotherapy & Hard & Intra-chromosomal, balanced:  & 10 & 19 & 1091 & 3449 \\
 &  & (frequency $\downarrow$ with age) &  &  &  & Pericentric inversion  &  &  &  &  \\
\midrule
inv(3)(q21q26) & EVI1 & 1-2\% AML & Poor & Chemotherapy; HSCT & Moderate & Intra-chromosomal, balanced & 16 & 23 & 1097 & 3469 \\
 & rearrangement & $<1 \%$ MDS &  &  &  & Paracentric inversion &  &  &  &  \\
\midrule
t(11;19)(q23.3;p13.1) & KMT2A::ELL & $<1\%$ Myeloid neoplasm & Poor & Chemotherapy; HSCT & Moderate & Inter-chromosomal & 14 & 47 & 1095 & 3497 \\
t(11;19)(q23;p13.3) & KMT2A::MLLT1 &  &  &  &  &  &  &  &  &  \\
\midrule
t(15;17)(q24;q21)\textsuperscript{4} & PML::RARA & 5-8\% AML & Good & ATRA-based therapy & Easy & Inter-chromosomal & 3 & 3 & 1081 & 3418 \\
\midrule
t(8;21)(q21;q22)\textsuperscript{4}& RUNX1::RUNX1T1 & 1-5\% AML & Good & Chemotherapy & Easy & Inter-chromosomal & 6 & 15 & 1081 & 3441 \\
\midrule
t(6;9)(p23;q34.1)\textsuperscript{4} & DEK::NUP214 & $<2\%$ AML\textsuperscript{5} & Poor & Chemotherapy, HSCT & Hard & Inter-chromosomal & 2 & 2 & 1081 & 3409 \\
&  & $<1\%$ Other myeloid neoplasm &  &  &  &  & & &  &  \\
\bottomrule
\end{tabular}
\caption{ \textbf{Summary of different chromosome aberrations evaluated in dataset.} \label{table:aberrations}
\emph{Abbreviations:}
AML, acute myeloid leukemia; APL, acute promyelocytic leukemia; ATRA, all-trans retinoic acid; B-ALL, B-lymphoblastic leukemia/lymphoma; CML, chronic myeloid leukemia; HSCT, hematopoietic stem cell transplantation; MDS, myelodysplastic syndromes; MPAL, mixed phenotype acute leukemia; TKI, Tyrosine kinase inhibitors.
\newline \textsuperscript{1} `Balanced' refers to an aberration being apparently balanced during microscopic assessment.
\newline  \textsuperscript{2} Depending on concurrent abnormalities.
\newline  \textsuperscript{3} One patient with t(16;16) was not included in the analysis data set for inv(16).
\newline  \textsuperscript{4} These translocations were seen in fewer than 10 patients and therefore not formally analysed in this study.
\newline  \textsuperscript{5} Also known as APL.
}
\end{sidewaystable}

\subsection{Low-frequency anomalies and \emph{de novo} detection} \label{sec:rare_anomalies_result}
\begin{figure}[htbp]
    \centering
    \begin{subfigure}{0.45\textwidth}
        \centering
        \includegraphics[width=\linewidth]{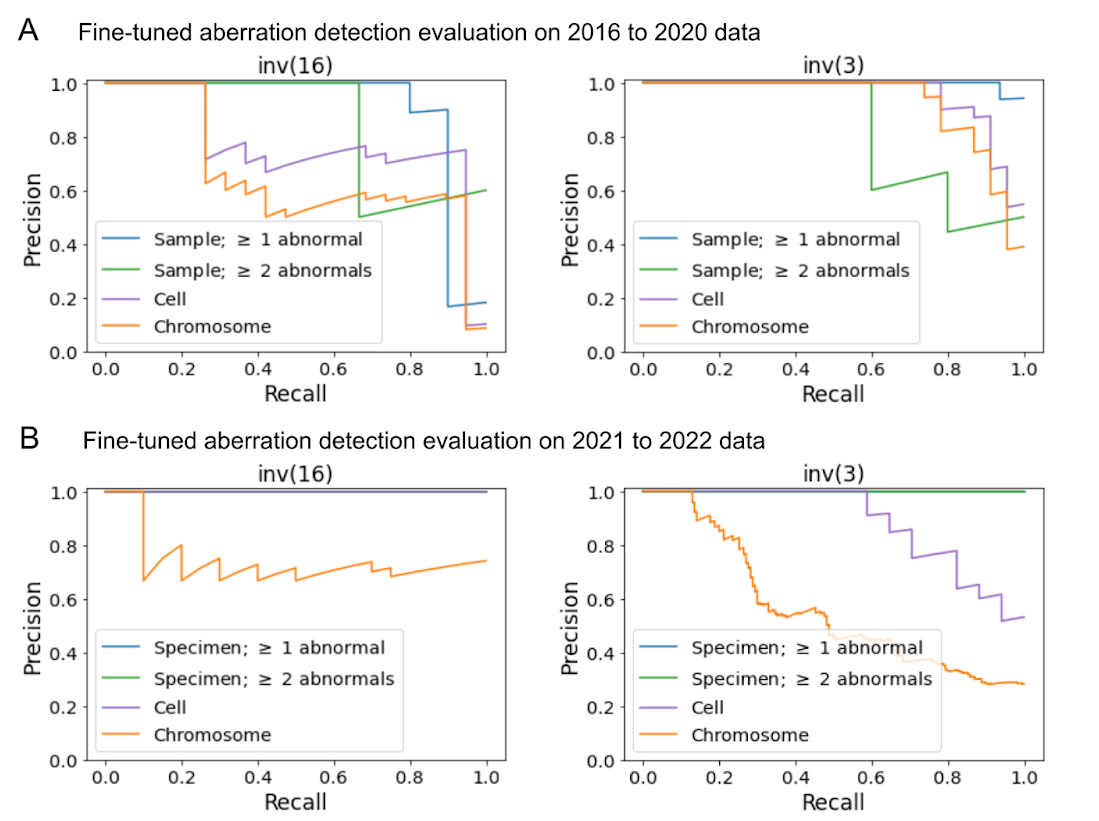}
        \caption{\emph{Fine-tuned}}
        \label{fig:finetuning_inv}
    \end{subfigure}
    \hfill
    \begin{subfigure}{0.45\textwidth}
        \centering
        \includegraphics[width=\linewidth]{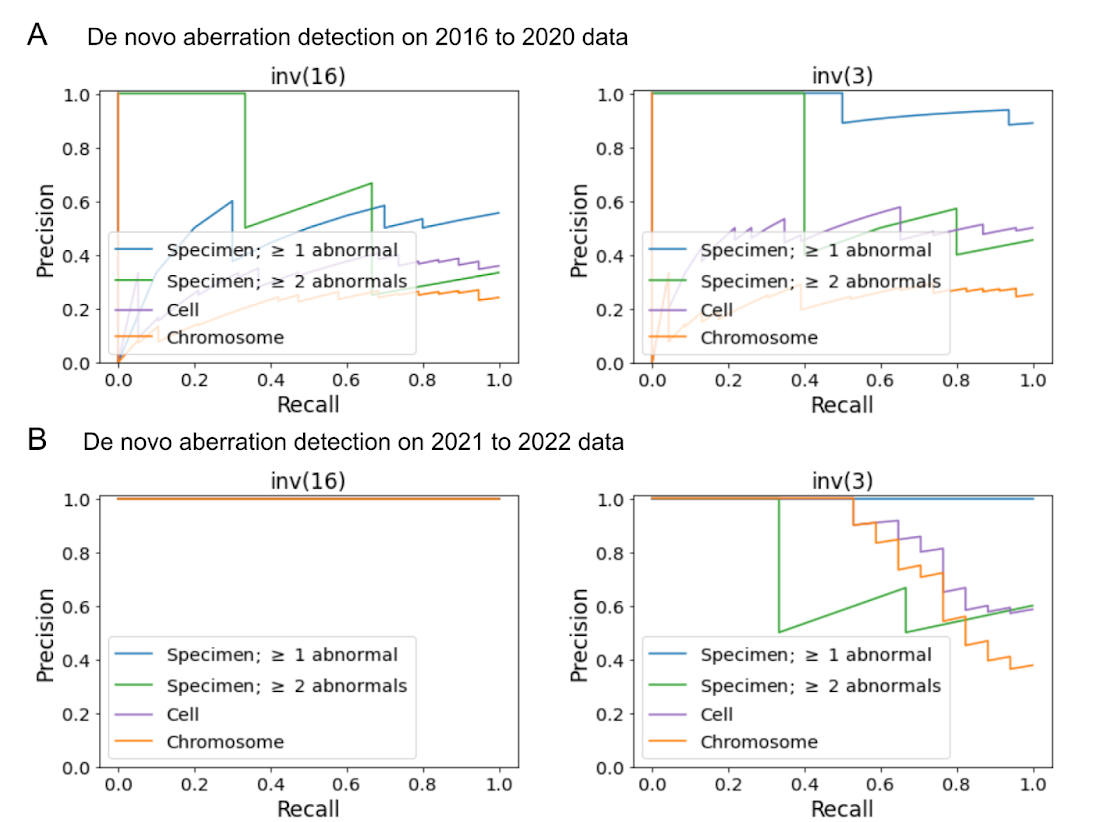}
        \caption{\emph{De-novo}}
        \label{fig:denovo_inv}
    \end{subfigure}
    \caption{\textbf{Comparison of fine-tuned and de novo models in detecting inversions 16 and 3.} 
    The performance of both approaches is compared across two validation datasets: 2016-2020 and 2021-2022. 
    Detection performance is shown at specimen ($\geq$ 1 abnormal, $\geq$ 2 abnormals), cell, and chromosome levels.}
    \label{fig:combined_inv}
\end{figure}

Fig.~\ref{fig:anomalies_2021_2022} shows model performance for rare aberrations, including \emph{de novo} detection, for the 2021-2022 validation dataset (analogous to Fig.~\ref{fig:rare_anomalies_result}).
Fig.~\ref{fig:combined_inv} shows results for the rare, challenging inv(16) and inv(3) anomalies. 


\begin{figure}\label{fig:anomalies_2021_2022}
    \centering
    \includegraphics[width=\linewidth]{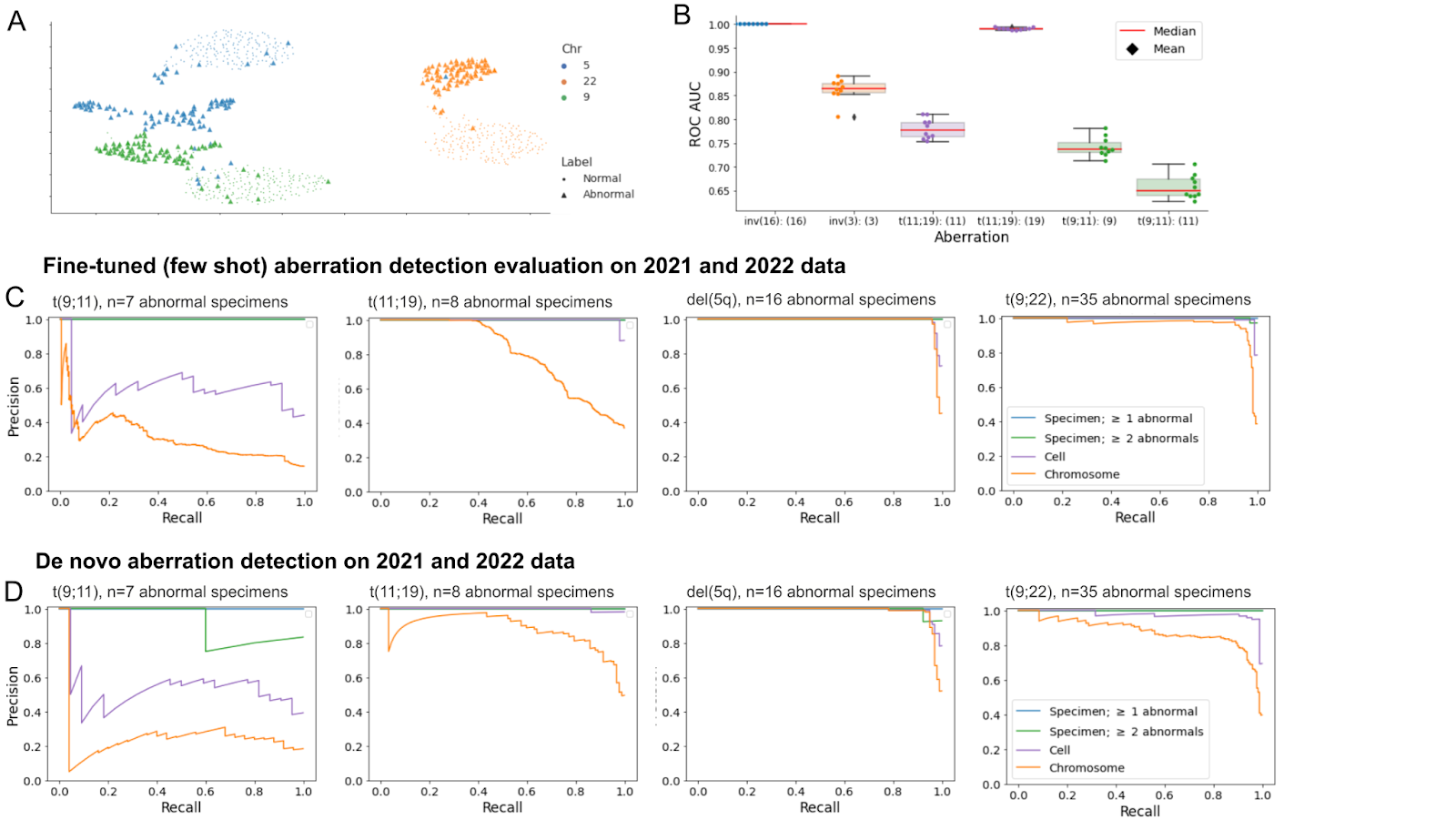}
    \caption{\textbf{Independent validation of model performance for rare aberrations using specimen-level precision-recall curves on a separate 2021-2022 clinical dataset.} 
    Counterpart to Figs \ref{fig:abb_pred} and \ref{fig:rare_anomalies_result} for the 2021-2022 validation dataset. 
    \emph{(A)} UMAP visualisation of normal and abnormal chromosomes 5, 9 and 22 from 2021-2022 dataset, using chromosome identification model trained on 2016-2020 dataset.
    \emph{(B)} ROC AUC for rare abnormalities in the 2021-2022 dataset, evaluated on 10 models fine-tuned on training sets from 2016-2020 data. 
    \emph{(C)} Precision-recall curves for t(9;11), t(11;19), del(5q), and t(9;22), at the individual chromosome image level (orange) and aggregated at the cell (purple) or specimen levels. For specimen level $\geq 1$ abnormal (blue) the single highest probability abnormal chromosome was used, for specimen level $\geq 2$ abnormal (green) the second highest probability abnormal chromosome was used. 
    \emph{(D)} Precision-recall for \emph{de novo} aberration detection based on distance to N-nearest point (here 50th) from chromosome identification model.
    }
    \label{fig:anomalies_2021_2022}
\end{figure}

\subsection{Entropy filtering and challenging examples}
\begin{figure}
    \centering
    \includegraphics[width=0.6\linewidth]{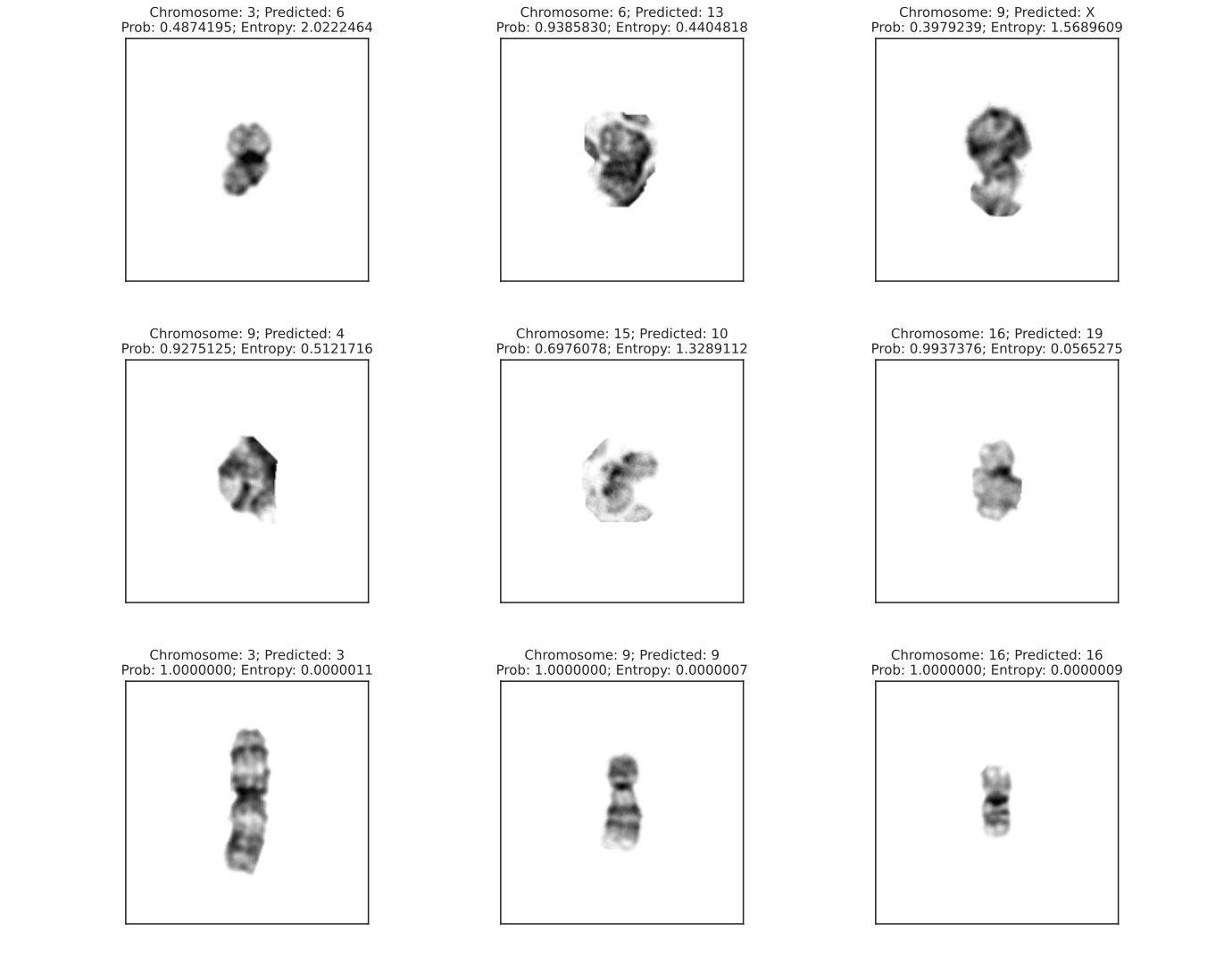}
    \caption{\textbf{Examples of chromosome images illustrating the effect of image quality on the identification performance.} The predicted and true labels are shown above each image, along with the model's probability and prediction entropy. Top and middle rows depict examples where the model makes errors due to poor-quality chromosome images, which are characterised by high entropy.  
    In contrast, the bottom row shows examples of good-quality chromosomes with low entropy. 
    The model is confident in these predictions, and the predicted label matches the true label.
    }\label{fig:challenging_examples}
\end{figure}

Fig.~\ref{fig:filtered_confusion} shows an updated chromosome identification confusion matrix, as in panel B of Fig.~\ref{fig:chrmid-fig}, after removal of high-entropy examples. 
It is even closer to diagonal with very few discordant predictions. 
Fig.~\ref{fig:challenging_examples} gives examples of chromosomes for which entropy is high, characterised by overlaps, poor morphology and cellular debris.
We also show counterexamples for which chromosomes are clear so the model is confident.

\begin{figure}\label{fig:filtered_confusion}
    \centering
    \includegraphics[width=0.6\linewidth]{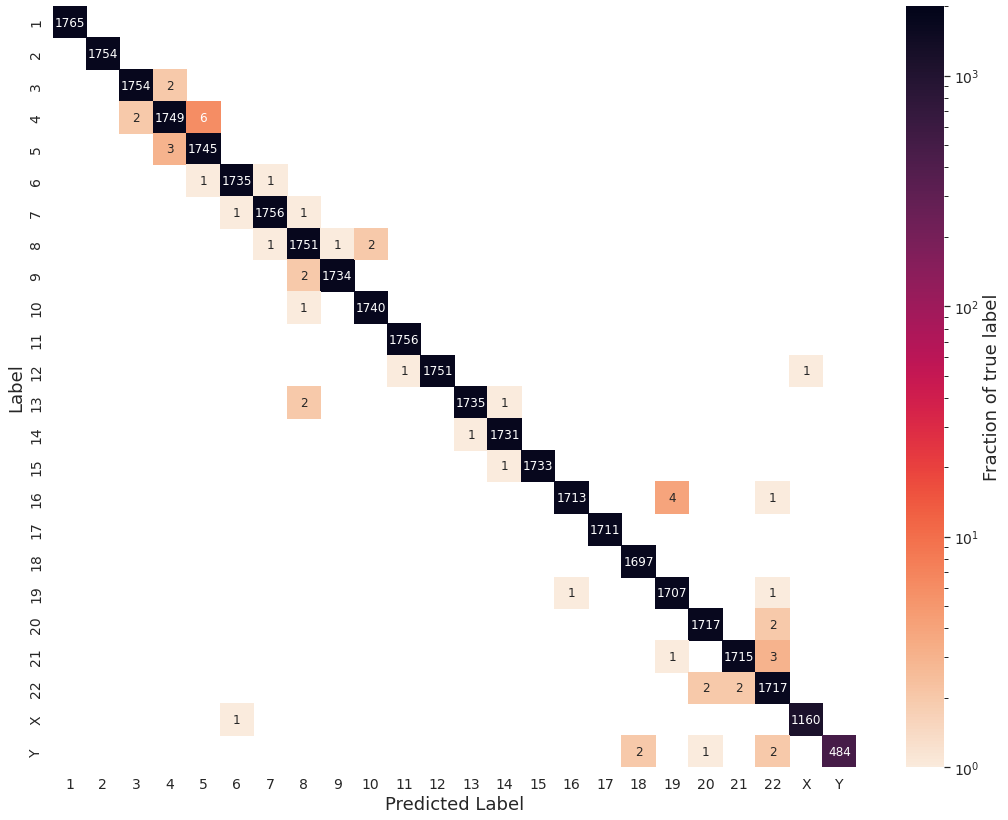}
    \caption{\textbf{Confusion matrix for chromosome identification, with entropy-filtered predictions.} This figure shows the confusion matrix of chromosome identification model, but after the predictions with high entropy were filtered, using an entropy cutoff at $10^{-5}$. The horizontal axis denotes the predicted labels and the vertical axis denotes the true labels. The figure shows the number of correct predictions along the diagonal, and the specific mistakes for each chromosome are visible as off-diagonal elements.
    }
    \label{fig:filtered_confusion}
\end{figure}

\end{document}

%% file: pipeline_schematic.tex
\def\offset{4.25}

\begin{tikzpicture}
\begin{pgfonlayer}{background}
    \foreach \i in {3, 2, 1} {
        \draw[black, thick, rounded corners=3pt, fill=white, fill opacity=1.0]
          (-1.5+\i*0.2, 2.5-\i*0.2) rectangle (15+\i*0.2, -4.5-\i*0.2);
    }
\end{pgfonlayer}

    \node(leftimage) at (0,0) {\includegraphics[width=2cm]{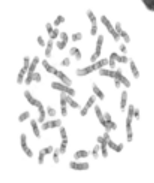}};
    \node(rightimage) at (5,0) {\includegraphics[width=2cm]{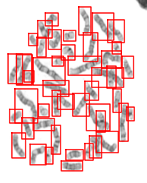}}; 
    \node[draw=black, fill=white, text=black, minimum height=1cm, 
           rounded corners=5pt, thick, align=center](owlvit) 
    at (2.5,0) {OWL-ViT};
    \node[](chromosomes) 
    at (2.5,1.75) {text prompt: `\texttt{chromosomes}'};
    \node[] 
    at (0,-1.5) {\texttt{metaphase}};
    \node[]() 
    at (0,-1.9) {\texttt{image}};
    \node[font=\Large] 
    at (-0.5,-4.1) {$\times N$};
    \draw[->, thick] (0.9,0) -- (owlvit);
    \draw[->, thick] (owlvit) -- (4.0,0);
     \draw[->, thick] (chromosomes) -- (owlvit);

     \node(single_chrm) at (7,0) {\includegraphics[width=1.0cm]{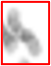}}; 
     \node(single_chrm_segd) at (10,0) {\includegraphics[width=1.0cm]{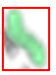}}; 

      \node[draw=black, fill=white, text=black, minimum height=1cm, 
           rounded corners=5pt, thick, align=center](sam) 
    at (8.5,0) {SAM};

    \draw[->, thick] (5.9,0) -- (6.53,0);
    \draw[->, thick] (7.44,0)-- (sam);
    \draw[->, thick] (sam) -- (9.54,0);

    \node[draw=black, fill=white, text=black, minimum height=1cm, 
           rounded corners=5pt, thick, align=center ](chrmidvit) 
    at (12,0) {chrmID ViT};

    \foreach \i in {1, 2, ..., 9} {
        \node[draw, circle, font=\small, minimum size=0.45cm, inner sep=0pt, thick] (n\i) at (\offset + 0.7*\i, -2) {\i};}

    \foreach \i in {20,21,22} {
        \node[draw, circle, font=\small, minimum size=0.45cm, inner sep=0pt, thick] (n\i) at (\offset + 0.7*\i - 9*0.7, -2) {\i}; 
        }
    \node[font=\small, minimum size=0.45cm, inner sep=0pt] (dots) at (\offset + 0.7*10, -2) {$\cdots$};
    \node[draw, circle, font=\small, minimum size=0.45cm, inner sep=0pt, thick] (nX) at (\offset + 0.7*14, -2) {X};
     \node[draw, circle, font=\small, minimum size=0.45cm, inner sep=0pt, thick] (nY) at (\offset + 0.7*15, -2) {Y};

     \draw[->, thick] (10.4,0) -- (chrmidvit);
     \draw[ thick] (11.05,-0.43) -- (n5);
     \draw[ thick] (chrmidvit) -- (n9);
     \draw[ thick] (chrmidvit) -- (n22);

    \node[draw=black, fill=white, text=black, minimum height=1cm, 
       rounded corners=5pt, thick, align=center](del5vit) at (\offset + 0.7*5, -3.5) {del(5q) ViT};
    \node[draw=black, fill=white, text=black, minimum height=1cm, 
       rounded corners=5pt, thick, align=center](t922-9) at (\offset + 0.7*9 , -3.5) {t(9;22)-9 ViT};
    \node[draw=black, fill=white, text=black, minimum height=1cm, 
       rounded corners=5pt, thick, align=center](t922-22) at (\offset + 0.7*13 , -3.5) {t(9;22)-22 ViT};

     \draw[->, thick] (n5) -- (del5vit);
     \draw[->, thick] (n9) -- (t922-9);
     \draw[->, thick] (n22) -- (t922-22);
        
    \draw[thick] (\offset + 0.7*5, -4.0) -- (\offset + 0.7*5, -5.5) ;
    \draw[thick] (\offset + 0.7*9, -4.0) -- (\offset + 0.7*9, -5.5) ;
    \draw[thick] (\offset + 0.7*13, -4.0) -- (\offset + 0.7*13, -5.5) ;

    \draw[thick] (\offset + 0.7*5+0.2, -4.7) -- (\offset + 0.7*5+0.2, -5.5) ;
    \draw[thick] (\offset + 0.7*9+0.2, -4.7) -- (\offset + 0.7*9+0.2, -5.5) ;
    \draw[thick] (\offset + 0.7*13+0.2, -4.7) -- (\offset + 0.7*13+0.2, -5.5) ;
    
    \draw[thick] (\offset + 0.7*5+0.4, -4.9) -- (\offset + 0.7*5+0.4, -5.5) ;
    \draw[thick] (\offset + 0.7*9+0.4, -4.9) -- (\offset + 0.7*9+0.4, -5.5) ;
    \draw[thick] (\offset + 0.7*13+0.4, -4.9) -- (\offset + 0.7*13+0.4, -5.5) ;

     \node[draw=black, fill=white, text=black, minimum height=1cm, 
       rounded corners=5pt, thick, align=center](aggregate) at (\offset + 0.7*9 , -6) {Aggregrate $N\geq 20$ metaphases for \textbf{patient-level} prediction};

\end{tikzpicture}